\documentclass[9pt, conference]{IEEEtran}
\IEEEoverridecommandlockouts
\usepackage{fontawesome5}
\usepackage{cite}
\usepackage{array}
\usepackage{booktabs} 
\usepackage{amsmath,amssymb,amsfonts}
\usepackage[linesnumbered,ruled,vlined]{algorithm2e}
\usepackage{amsmath}

\usepackage{algpseudocode}
\usepackage{tikz}
\usepackage{verbatimbox} 
\usepackage{graphicx}
\usepackage{textcomp}
\usepackage{xcolor}
\usepackage{listings}
\usepackage{multirow}
\usepackage{caption}
\usepackage{pifont}
\usepackage{graphicx}
\usepackage{booktabs}
\usepackage{colortbl}
\usepackage{xcolor}
\usepackage{array}
\usepackage{multirow}
\usepackage{hyperref}
\usepackage{makecell}
\usepackage[a4paper, total={184mm,239mm}]{geometry}
\def\BibTeX{{\rm B\kern-.05em{\sc i\kern-.025em b}\kern-.08em
    T\kern-.1667em\lower.7ex\hbox{E}\kern-.125emX}}
    
\definecolor{keywordcolor}{rgb}{0, 0, 0}
\definecolor{commentcolor}{rgb}{0.5, 0.5, 0.5}
\definecolor{stringcolor}{rgb}{0.5, 0.5, 0.5}
\definecolor{backgroundcolor}{rgb}{0.97, 0.97, 0.97} 
\lstdefinestyle{customverilog}{
  language=Verilog,
  basicstyle=\ttfamily\footnotesize,
  keywordstyle=\color{keywordcolor}\bfseries,
  commentstyle=\color{commentcolor},
  stringstyle=\color{stringcolor},
  morekeywords={module, if, else, case, end, begin, endcase, input},
  showstringspaces=false,
  numberstyle=\tiny\color{red},
  frame=single,
  rulecolor=\color{black},
  backgroundcolor=\color{white},
  tabsize=2,
  captionpos=b,
  breaklines=true,
  postbreak=\mbox{\textcolor{red}{$\hookrightarrow$}\space},
}

\usepackage[most]{tcolorbox}
\usepackage{listings}
\usepackage{xcolor}
\usepackage{enumitem}

\newtcolorbox{verificationbox}[2][]{
    enhanced,
    colframe=#2!75!black,
    colback=#2!3!white,
    arc=2mm,
    boxrule=0.8pt,
    left=3mm, right=3mm, top=2mm, bottom=2mm,
    drop shadow={shadow xshift=0.3mm, shadow yshift=-0.3mm, opacity=0.2},
    title={#1}, 
    fonttitle=\bfseries\small,
    coltitle=#2!90!black,
    attach boxed title to top left={xshift=0.5cm, yshift=-2mm},
    boxed title style={
        size=small,
        colback=#2!20!white,
        colframe=#2!70!black,
        arc=1.5mm
    },
    overlay first={
        \node[#2!75!black] at ([xshift=5mm]frame.north west) 
            {\scriptsize\faIcon[solid]{code}};
    }
}

\begin{document}

\title{AutoVeriFix: Automatically Correcting Errors and Enhancing Functional Correctness in LLM-Generated Verilog Code\\}
\author{\IEEEauthorblockN{Yan Tan, Xiangchen Meng, Zijun Jiang and Yangdi Lyu$^{\dagger}$}
\IEEEauthorblockA{Microelectronics Thrust, The Hong Kong University of Science and Technology (Guangzhou) \\
$^{\dagger}$Corresponding author: yangdilyu@hkust-gz.edu.cn}
}
\maketitle

\begin{abstract}
Large language models (LLMs) have demonstrated impressive capabilities in generating software code for high-level programming languages such as Python and C++. However, their application to hardware description languages, such as Verilog, is challenging due to the scarcity of high-quality training data. Current approaches to Verilog code generation using LLMs often focus on syntactic correctness, resulting in code with functional errors. To address these challenges, we present AutoVeriFix, a novel Python-assisted two-stage framework designed to enhance the functional correctness of LLM-generated Verilog code. In the first stage, LLMs are employed to generate high-level Python reference models that define the intended circuit behavior. In the second stage, these Python models facilitate the creation of automated tests that guide the generation of Verilog RTL implementations. Simulation discrepancies between the reference model and the Verilog code are iteratively used to identify and correct errors, thereby improving the functional accuracy and reliability of the LLM-generated Verilog code. Experimental results demonstrate that our approach significantly outperforms existing state-of-the-art methods in improving the functional correctness of generated Verilog code.
\end{abstract}


\section{Introduction}
Large language models (LLMs) have transformed software development by enhancing productivity through automated code generation. They translate natural language descriptions into high-quality code, enabling developers to focus on design and problem-solving rather than routine tasks. This boost in productivity stems from LLMs being trained on extensive open-source code, particularly in popular languages like Python and C++~\cite{Mastropaolo_Pascarella_Guglielmi_Ciniselli_Scalabrino_Oliveto_Bavota_2023, Nijkamp_Hayashi_Xiong_Savarese_Zhou}.

Building on these advancements, researchers are exploring the use of LLMs for generating hardware code. Initial studies indicate promise in creating basic structures for hardware description languages (HDLs)~\cite{Benchmark_RTL_2022, Dehaerne_Verilog_2023, OpenLLM, VerilogEval, Xie_2023, 2024origen}. However, challenges persist in matching the performance of software code generation~\cite{chen2021codex}. While HDLs share features with software languages, they require precise circuit behavior specification at the register-transfer level (RTL), making direct application of LLMs prone to errors. Additionally, the scarcity of high-quality RTL code datasets, compared to the abundance of open-source software, limits the training potential for HDL-specific LLMs.

To reduce the ratio of buggy designs generated by LLMs, researchers have introduced self-reflection into their frameworks~\cite{AutoChip,2024origen,RTLFixer,verilogcoder} to identify syntax errors and refine outputs to ensure correctness and adherence to the design specifications. For example, OriGen~\cite{2024origen} proposes a code-to-code augmentation process with two datasets: an enhanced code dataset and an error correction dataset. However, existing methodologies primarily focus on syntactic corrections through pattern matching and template-based approaches, often neglecting functional errors. We have observed significant functional errors in the Verilog code produced by these approaches, particularly in common hardware design tasks such as multi-branch logic and complex state transitions. 

The cost of designing buggy hardware code is considerably higher than that of software code, as errors can lead to substantial expenses in the electronic design automation and manufacturing processes. To address the challenge of generating functionally correct Verilog code, we propose AutoVeriFix, a two-stage approach that integrates an LLM-assisted automated testing mechanism to fix errors in LLM-generated RTL code. In the first stage, an LLM generates a Python-based reference model from hardware specifications. The Python reference model is mostly functionally correct due to the strong capabilities of existing LLMs in understanding and generating Python code, allowing it to define the intended behavior of the circuit. \emph{Note that the generated Python code by LLMs cannot be directly converted into Verilog, as it tends to be too complex for high-level synthesis (HLS) tools like MyHDL~\cite{myhdl} and SODA~\cite{soda}, which only support a specific synthesizable subset of Python.}
Next, expected input-output pairs (testbench) are generated based on simulation results from the reference model, using an iterative process that involves coverage analysis to refine the testbench for improved validation. In the second stage, a different LLM is utilized to generate the Verilog code, which is then corrected using the testbench from the first stage. The simulation results of the Verilog code are compared to those of the reference model to verify functional correctness. If discrepancies are found, i.e., mismatches in input-output pairs, this information is fed back into the LLM-assisted RTL generation process, providing targeted guidance for iterative improvements in code generation. While this approach cannot guarantee 100\% functional correctness, experimental results demonstrate that this two-stage method automatically corrects a large portion of errors in the LLM-generated Verilog code.
 
Our main contributions are summarized as follows.
\begin{enumerate}
    \item We propose a novel framework integrating an LLM-assisted automated testing mechanism with LLM-assisted hardware code generation~\footnote{We will open source our code after the paper is accepted.}. This design leverages the capability of LLMs to accurately generate high-level software languages to define the desired circuit behavior and guide automatic Verilog generation. 
    
    \item  We introduce a novel approach for generating a reference model using LLMs from hardware specifications. This method not only produces high-level software code but also develops a high-coverage testbench. 
    
    \item We propose an LLM-based RTL code generation that incorporates feedback from testbench discrepancies. This iterative and feedback-driven RTL code generation process improves both the syntactic and functional correctness of the generated code.
    
    \item Our experimental results show a significant improvement over existing state-of-the-art methods in achieving higher functional correctness for RTL code generation, with our approach achieving over 80\% correctness on current benchmarks.
    
\end{enumerate}

\section{BACKGROUND AND PRIOR WORK}




\begin{figure*}
    \centering
    \includegraphics[width=0.8\linewidth]{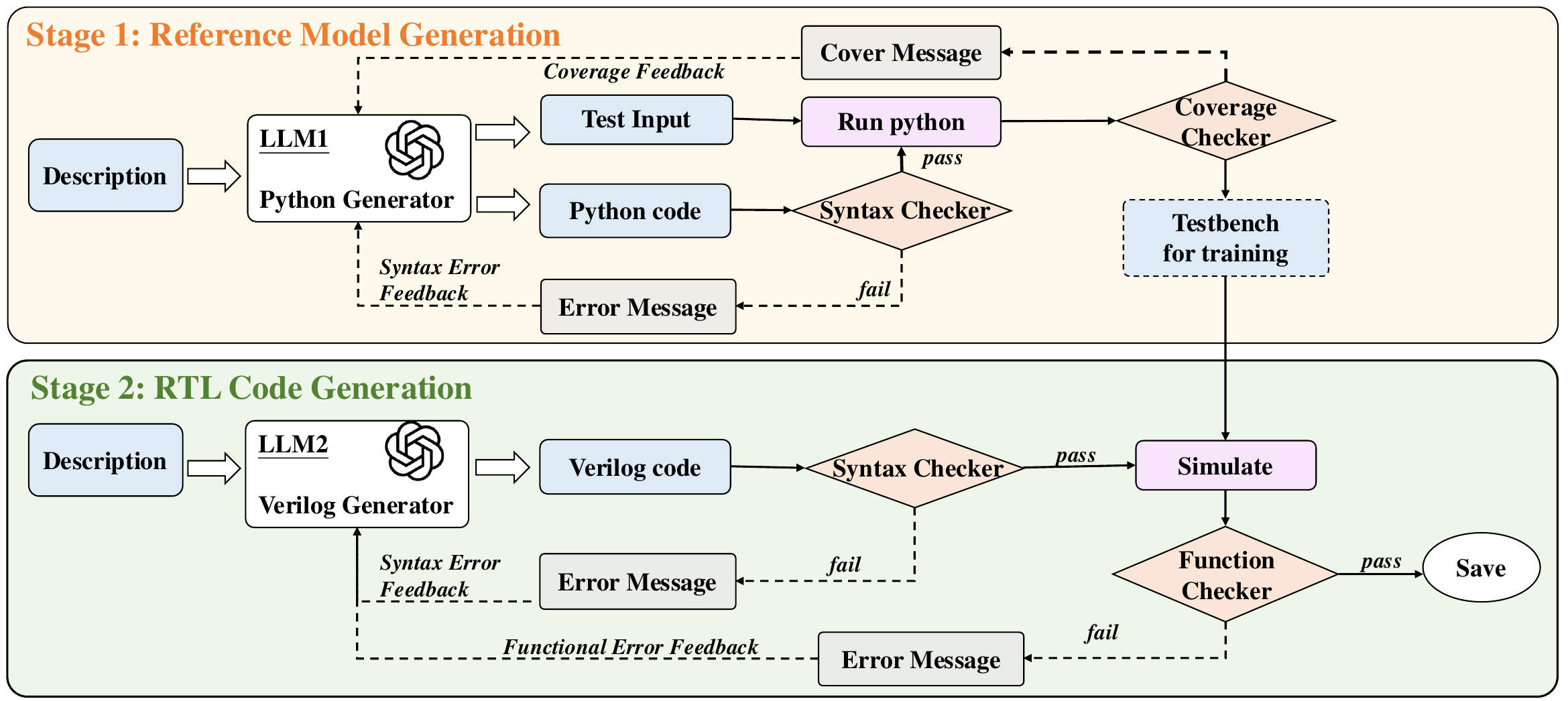}
    \caption{The two-stage framework of AutoVeriFix. In the first stage, an LLM generates a reference model in Python along with a testbench (input-output pairs) based on the hardware description, while the second stage utilizes another LLM to produce the Verilog code and iteratively correct any errors using the testbench.}
    \vspace{-0.1in}
    \label{fig:overview}
\end{figure*}
While HDLs share many common features with software languages, applying LLMs to hardware design generation poses unique challenges, particularly the need for precise specifications of timing, logic, and signal behaviors. Additionally, the limited availability of high-quality, publicly accessible RTL datasets restricts the ability of LLMs to train effectively on hardware-specific tasks~\cite{chen2021codex,BetterV}. 

To address these challenges, researchers have explored multiple strategies to improve the quality and diversity of training data, as well as enhance the correctness of generated RTL code. These strategies can be broadly categorized into data-centric approaches and feedback-driven mechanisms.

\subsection{Data-Centric Approaches}
Recent works have focused on curating hardware-specific datasets and augmenting them to improve model performance. Thakur \textit{et al.}~\cite{thakur2024verigen} constructed an unsupervised Verilog dataset sourced from GitHub repositories and Verilog textbooks to fine-tune CodeGen-16B~\cite{nijkamp2022codegen}. However, due to the absence of manual annotations in this dataset, models fine-tuned on this dataset still underperform compared to commercial tools such as GPT-3.5. In contrast, the NVIDIA research team~\cite{VerilogEval} developed a supervised dataset by carefully gathering, filtering, and refining Verilog files and design snippets from open-source hardware repositories. These rigorous steps ensure the inclusion of representative RTL code, enabling LLMs to learn hardware design patterns more effectively. Fine-tuned on the supervised dataset, VerilogEval~\cite{VerilogEval} became the first non-commercial model to match GPT-3.5 performance. Additionally, data augmentation techniques have been employed to further enhance dataset diversity. RTLCoder~\cite{Xie_2023} introduces a framework that generates natural language descriptions aligned with Verilog code semantics, creating contextual pairings that enrich the training process. Similarly, OriGen~\cite{2024origen} employs a code-to-code augmentation strategy, producing enhanced datasets that include both original designs and error-corrected versions of RTL code. While these approaches improve LLMs' understanding of RTL design principles, they are insufficient for ensuring syntax and functional correctness, as the generated code often fails to meet the rigorous verification standards required in hardware design.

\subsection{Feedback-driven Mechanism}
Feedback-driven mechanisms have been integrated into the code generation workflow to improve syntactic correctness in LLM-generated RTL code. Inspired by hardware design verification workflows, AutoChip~\cite{AutoChip} incorporates compilation error messages into the generation loop, enabling the model to identify and rectify syntax errors. RTLFixer~\cite{RTLFixer} uses Retrieval-Augmented Generation (RAG) and prompting techniques to address errors. It retrieves guidance from a database related to the compilation error and incorporates interactive debugging to fix syntax errors in the generated code. OriGen~\cite{2024origen} autonomously corrects syntax errors by leveraging compiler feedback, which enhances the accuracy and reliability of its code generation process. RTLCoder~\cite{Xie_2023} introduces a code quality scoring feedback mechanism, where the generated code is evaluated based on syntactic correctness and similarity to reference code. The scoring results are then used as feedback to enhance the model's generation capability. Although these methods improve the syntactic validity of generated Verilog code, they fall short in addressing functional correctness.



\section{METHODOLOGY}

\subsection{Overview}

To improve functional correctness, we propose AutoVeriFix, a two-stage feedback framework that automatically corrects errors in Verilog code generated by LLMs. The key idea is to leverage LLMs to generate both high-level programming code (Python) and hardware description language code (Verilog). Given that LLMs are extensively trained to produce Python code, the functional correctness of this code is nearly assured. As demonstrated in our experiments (Section~\ref{subsec:refeval}), the functional correctness of generated Python code in our benchmarks exceeds 95\%, while the functional correctness of the generated Verilog code using the same LLM model is less than 50\%. In the ideal case, using the generated Python code as an almost-perfect reference model can help boost the functional correctness of LLM-generated Verilog code close to that of the reference model. 

The two-stage framework of AutoVeriFix is shown in Fig.~\ref{fig:overview}. In the first stage, an LLM generates a reference model in Python along with a testbench (input-output pairs) derived from the hardware description. The second stage involves another LLM generating the Verilog code and employing the testbench generated from the first stage to iteratively correct any errors found in the generated Verilog code. The quality of the Verilog code generated in this second stage is influenced by the testbench produced in the first stage. To enhance this quality, we adopt a feedback-driven approach to improve the coverage of the testbench, where the LLM uses coverage metrics and information about uncovered branches to refine the generated tests. As shown in our experiments (Section~\ref{subsec:testeffect}), the testbench generated by this iterative refinement process significantly boosts the correctness of the generated Verilog compared to the original testbench created by the LLM.

In summary, our method innovatively leverages high-level language models to guide hardware code generation and explicitly incorporates high-level language simulation coverage as a reference for verifying hardware functionality. This framework effectively addresses the shortcomings of existing LLM-based hardware generation, particularly the absence of adaptive feedback from the functional verification process. Consequently, our approach significantly reduces the risks associated with uncovered hardware logic bugs, thereby improving the reliability of hardware designs generated automatically by LLMs.

\subsection{Reference Model Generation}
In the first stage, our framework provides the LLM with a structured functional description of the hardware design along with the task of generating a Python reference model and an initial set of tests. Unlike HLS tools that translate high-level Python or C++ into Verilog, our framework uses Python code solely as a functional oracle. This reference model is generated from natural language and used to validate and correct Verilog code via simulation. 

\begin{figure}[t]
    \centering
    \includegraphics[width=0.95\linewidth]{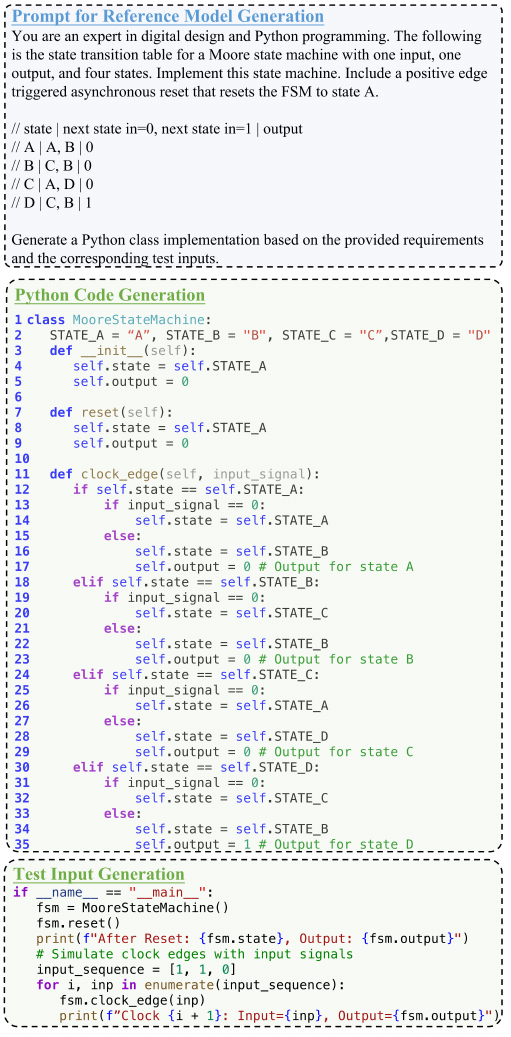}
    \caption{Our framework provides the LLM with a structured functional description of the hardware design, and the LLM generates the Python class and the initial test sequence.}
    \label{fig:reference}
\end{figure}

In the simple example illustrated in Fig.~\ref{fig:reference}, we start by giving the LLM a structured description of a state machine (shown at the top of Fig.~\ref{fig:reference}). The LLM then generates a Python class that captures the behavior of the state machine (illustrated in the middle of Fig.~\ref{fig:reference}) and an initial input sequence of [1, 1, 0] (displayed at the bottom of Fig.~\ref{fig:reference}). Once the Python class and test inputs are created, they undergo a syntax check. If syntax errors are detected, the framework provides feedback in the form of error messages, allowing the LLM to adjust its output as necessary, as indicated in the top part of Fig.~\ref{fig:overview}. Once the syntax is verified, the Python model is executed using the generated test inputs, and the simulation results—including input, state, and output at each time step—are recorded for future refinement of the testbench. 

In the example provided in Fig.~\ref{fig:reference}, the generated reference model effectively captures the functional behavior of the hardware. This success can be attributed to the extensive training of LLMs on large Python datasets, which enables them to produce highly accurate and readable Python implementations that align with the functional descriptions given. This Python-based reference design serves as the foundation for guiding the hardware generation and verification process.

\subsection{Testbench Refinement}
\begin{figure}[h]
    \centering
    \includegraphics[width=1\linewidth]{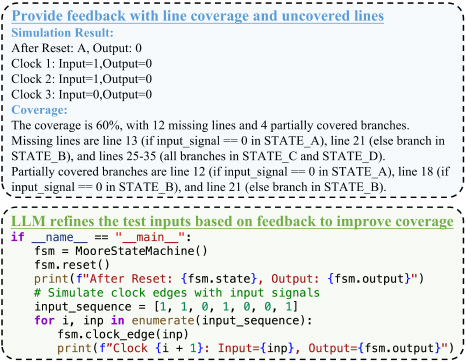}
    \caption{LLM refines the test inputs with coverage feedback}
    \label{fig:refine}
\end{figure}

While the generated reference model is generally accurate, the initial test inputs often fall short of our expectations, typically resulting in low coverage. We simulate the Python code using these test cases and assess line coverage to determine which functional paths within the model have been exercised. To ensure sufficient testing, we set a minimum acceptable standard of 85\% for line coverage. If the measured line coverage is below this threshold, it indicates that the current test inputs are inadequate for thoroughly verifying the model. In such cases, we provide feedback that includes information on uncovered branches and coverage metrics to guide the LLM in generating new test inputs. For example, as illustrated in Fig.~\ref{fig:refine}, the coverage analysis shows that only 60\% of the code is covered, with 12 lines and 4 branches remaining uncovered. This feedback identifies specific areas of the model that need additional test cases for complete verification. The LLM then utilizes this feedback to refine the test inputs, particularly focusing on the uncovered branches, ensuring that all functional scenarios are tested. For example, additional input vectors like [1, 1, 0, 1, 0, 0, 1] are generated to exercise the uncovered cases in the Python model in this example. This iterative process continues until the desired coverage threshold is met.

Once the test inputs meet the coverage requirements, they are converted into a hardware testbench for Verilog code generation in the next stage. This testbench facilitates error correction and helps ensure, though not with absolute certainty, that the hardware implementation aligns with the original functional specifications.

\subsection{Verilog Generation}
Once a high-coverage testbench is established, we proceed to generate Verilog code from the same hardware description using a new LLM. We then conduct a syntax check to ensure that the generated Verilog code compiles correctly. As with the reference model generation, any error messages encountered are provided to the LLM for correction of syntax issues. Next, we apply the hardware testbench created in the first stage to verify the functional correctness of the generated Verilog code. If the simulation results match the testbench, the hardware code is accepted as a successful implementation. Otherwise, the simulation iteratively provides feedback, including errors and inconsistencies, which is then used to guide further rounds of Verilog code generation and optimization, as illustrated in Fig.~\ref{fig:overview}. Through this iterative feedback process, we ultimately ensure that the Verilog hardware code aligns with the testbench.

\subsubsection{Verilog Generation}
\begin{figure}
    \centering
    \includegraphics[width=0.95\linewidth]{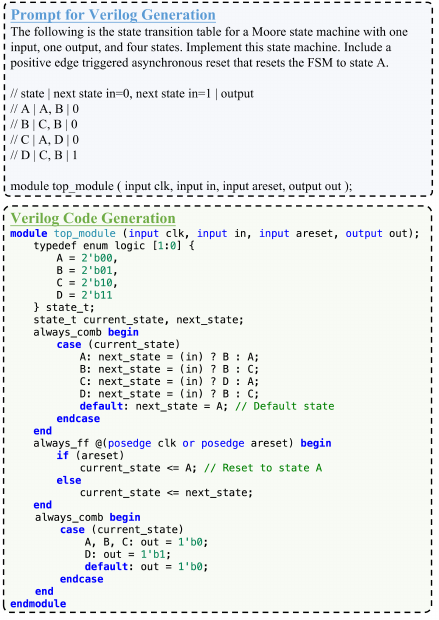}
    \caption{Verilog generation}
    \label{fig:verilog}
\end{figure}

In order to help the LLM generate the correct Verilog code, we provide the input hardware description in the prompt. This description includes module definitions, input/output ports, combinational and sequential logic, as well as state machine behavior, as shown at the top of Fig.~\ref{fig:verilog}. The LLM then generates a complete Verilog implementation that reflects the specified design requirements, as shown at the bottom of Fig.~\ref{fig:verilog}.

\subsubsection{Syntax Debugging}
Syntax errors, such as undeclared variables, missing semicolons, or misplaced keywords, can hinder successful code compilation. Syntax debugging aims to identify and resolve these issues by compiling the generated Verilog code using hardware simulation tools, which detect and report any syntax errors. The example of syntax debug prompt is shown at the top of Fig.~\ref{fig:debug}. The prompt includes detailed compile error messages, such as the type of error, its location in the code, and additional diagnostic information. This debug information, along with the original code, is fed back into the LLM, allowing it to analyze the errors and generate corrections.

The corrected code is then recompiled to verify that the errors have been addressed. If additional syntax issues arise, new error messages are generated and returned to the LLM. This iterative process continues until the Verilog code successfully passes all syntax checks, ensuring it is syntactically correct and ready for the next validation phase.

\subsubsection{Function Debugging}
\begin{figure}[t]
    \centering
    \includegraphics[width=0.95\linewidth]{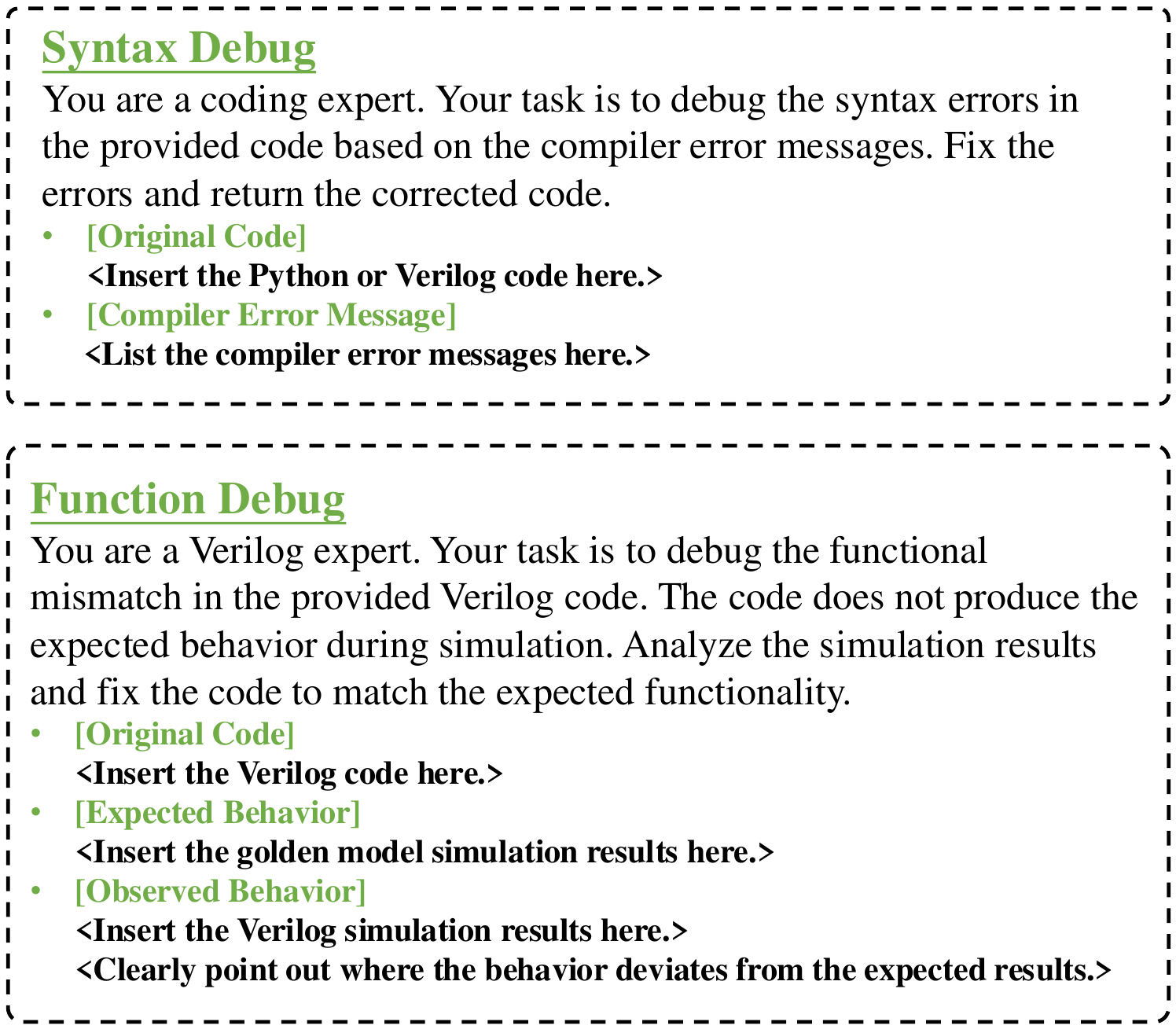}
    \caption{Syntax error and functional error feedback}
    \label{fig:debug}
\end{figure}

Once the Verilog code passes syntax checks, it undergoes functional evaluation to confirm that its behavior meets the design specifications for all test conditions specified in the testbench from the reference model generation stage. The testbench supplies input stimuli to the Verilog implementation along with the expected output from the reference model. The outputs observed from the simulation of the Verilog code are then compared to these expected results.

If any discrepancies arise between the simulation results, the functional debugging process identifies these mismatches and feeds the relevant information back to the LLM, including the original Verilog code, the expected behavior, and the observed behavior, as illustrated at the bottom of Fig.~\ref{fig:debug}. This feedback enables the LLM to refine the Verilog code to better align with the expected functionality. The updated code is then resimulated, and the outputs are reassessed. This iterative process continues until the observed behavior matches the expected results, ensuring functional consistency.

\section{Experiment}

\subsection{Experimental Setup}
We implemented AutoVeriFix using OpenAI's GPT-3.5~\cite{openai2023gpt35} and GPT-4~\cite{openai2023gpt4} as the LLM tools for code generation. These models demonstrate exceptional performance in handling Python-based problems, which makes them highly effective in creating reference models. 

To evaluate the performance of Verilog code generation, we selected two representative benchmarks: VerilogEval~\cite{VerilogEval} and RTLLM~\cite{lu2024rtllm}. These datasets were specifically designed to evaluate the functional correctness and quality of Verilog code generated by language models. The VerilogEval dataset was derived from Verilog programming tasks sourced from the HDLBits instructional website. These tasks were divided into two subsets: VerilogEval-Human, with 156 carefully hand-crafted problems, and VerilogEval-Machine, comprising 143 problems generated by GPT. The RTLLM dataset v1.1 contains a total of 29 designs, while RTLLM v2.0 offers 50 diverse Verilog tasks that closely resemble real-world RTL design scenarios, providing varying levels of difficulty to evaluate the performance of AutoVeriFix and other existing approaches.

\subsection{Reference Model Correctness and Coverage of Testbench}
\label{subsec:refeval}

To validate the effectiveness of the reference model, we first conducted an analysis of the generated Python code and the testbench on three metrics: syntactic correctness, functional correctness, and line coverage. Syntactic correctness assesses whether the generated code compiles successfully without syntax errors. Functional correctness evaluates whether the code produces the expected outputs upon execution. Line coverage measures the proportion of code lines that are exercised by the final refined test inputs.

\begin{table}[h!]
\renewcommand{\arraystretch}{1.5}
\centering
\caption{Evaluation of the reference model and the testbench generated in Stage 1 using GPT-4.}
\setlength{\tabcolsep}{1mm}{
\begin{tabular}{l|c|c|c}
\toprule
\multirow{3}{*}{\textbf{Dataset}} & \multicolumn{2}{c|}{\textbf{Python Reference Model}} & \textbf{Testbench} \\\cline{2-4}
  & \textbf{Syntactic} & \textbf{Functional} & \textbf{Line}\\
    & \textbf{correctness (\%)}& \textbf{correctness (\%)} &  \textbf{coverage (\%)} \\ \midrule
\textbf{VerilogEval-human}    & 99.35   & 96.15  & 95.89   \\ 
\textbf{VerilogEval-machine}   & 100.00   & 98.60  & 97.09 \\
\textbf{RTLLM v1.1}             & 100.00    & 96.55  & 95.68     \\ 
\textbf{RTLLM v2.0}              & 98.00    & 94.00  & 91.16   \\ 
\bottomrule
\end{tabular}
\label{tab:reference_model_results}
} 
\end{table}

\begin{table*}[htbp]
\caption{Comparison of functional correctness on the VerilogEval benchmarks~\cite{VerilogEval} and the RTLLM benchmarks~\cite{lu2024rtllm}. The top scores ranked 1\textsuperscript{st}, 2\textsuperscript{nd}, and 3\textsuperscript{rd} in each column are highlighted in \colorbox[HTML]{FDE9E9}{Red}, \colorbox[HTML]{E0F7FA}{Blue}, and \colorbox[HTML]{E8F5E9}{Green}, respectively.}
\label{tab:comparison}
\centering
\renewcommand{\arraystretch}{1.5}
\setlength{\tabcolsep}{4pt}
\begin{tabular}{p{2.5cm}|p{4.5cm}|ccc|ccc|c|c}
\bottomrule
\rowcolor[HTML]{FFFFFF} 

{\multirow{3}{*}{\textbf{Category}}} & \multirow{3}{*}{\textbf{Model}} & \multicolumn{3}{c|}{\textbf{VerilogEval-human}} & \multicolumn{3}{c|}{\textbf{VerilogEval-machine}} & \textbf{RTLLM} & \textbf{RTLLM} \\

 &  &  \multicolumn{3}{c|}{\textbf{(\%)}} & \multicolumn{3}{c|}{\textbf{(\%)}} & \textbf{~v1.1 (\%)~} & \textbf{~v2.0 (\%)~} \\ \cline{3-10}

&  & \textbf{{pass@1}} & \textbf{pass@5} & \textbf{pass@10} & \textbf{{pass@1}} & \textbf{pass@5} & \textbf{pass@10} & \textbf{pass@5} & \textbf{pass@5} \\ \hline

\multirow{5}{2.5cm}{\textbf{Verilog-Specific Models}} 
 & VerilogEval~\cite{VerilogEval} & 28.8 & 45.9 & 52.3 & 46.2 & 67.3 & 73.7 & - & -\\  
   & CodeGen-6B MEV-LLM~\cite{nadimi2024multi} & 42.9 & 48.0 & 54.4 & 57.3 & 61.5 & 66.4 & - & -\\ 
& BetterV-CodeQwen~\cite{BetterV} & 46.1 & 53.7 & 58.2 & 68.1 & 79.4 & 84.5 & - & -\\ 
  & RTLCoder~\cite{Xie_2023} & 41.6 & 50.1 & 53.4 & 61.2 & 76.5 & 81.8 & 48.3 & - \\ 
 & OriGen~\cite{2024origen} & \cellcolor[HTML]{E8F5E9}54.4 & 60.1 & 64.2 & \cellcolor[HTML]{E0F7FA}74.1 & \cellcolor[HTML]{E0F7FA}82.4 & \cellcolor[HTML]{E0F7FA}85.7 & \cellcolor[HTML]{E8F5E9}65.5 & - \\ \hline

\multirow{3}{*}{\textbf{Open Source Models}} 
 & CodeLlama-7B-Instruct~\cite{roziere2023code} & 18.2 & 22.7 & 24.3 & 43.1 & 47.1 & 47.7 & 34.5 & 33.1\\ 
& CodeQwen1.5-7B-Chat~\cite{bai2023qwen} & 22.4 & 41.1 & 46.2 & 45.1 & 70.2 & 77.6 & 37.9 & 36.4 \\ 
& DeepSeek-Coder-7B-Instruct-v1.5~\cite{guo2024deepseek} & 31.7 & 42.8 & 46.8 & 55.7 & 73.9 & 77.6 & 37.9 & 36.4\\ \hline

\multirow{5}{*}{\textbf{Commercial LLM}} 
& Claude3-Sonnet~\cite{claude3_family} & 46.1 & 56 & 60.3 & 58.4 & 71.8 & 74.8 & 58.6 & 54.4 \\
& GPT-3.5~\cite{openai2023gpt35} & 35.6 & 48.8 & 52.6 & 49.4 & 72.7 & 77.6 & 44.8 & 36.2\\ 
& GPT-4~\cite{openai2023gpt4} & 43.5 & 55.8 & 58.9 & 60 & 70.6 & 73.5 & \cellcolor[HTML]{E8F5E9}65.5  & 58.7\\ 
& GPT-4 Turbo~\cite{openai2023gpt4} & 54.2 & \cellcolor[HTML]{E8F5E9}68.5 & \cellcolor[HTML]{E8F5E9}72.4 & 58.6 & 71.9 & 76.2 & \cellcolor[HTML]{E8F5E9}65.5 & \cellcolor[HTML]{E8F5E9}63.4\\  \hline
 
\multirow{2}{*}{\textbf{Ours}} 
 & AutoVeriFix with GPT-3.5 & \cellcolor[HTML]{E0F7FA}58.5 & \cellcolor[HTML]{E0F7FA}71.8 & \cellcolor[HTML]{E0F7FA}73.7 & \cellcolor[HTML]{E8F5E9}69.5 & \cellcolor[HTML]{E8F5E9}77.6 & \cellcolor[HTML]{E8F5E9}79.7 & \cellcolor[HTML]{E0F7FA}75.9 & \cellcolor[HTML]{E0F7FA}71.9\\ 
& AutoVeriFix with GPT-4 & \cellcolor[HTML]{FDE9E9}77.2 & \cellcolor[HTML]{FDE9E9}82.7 & \cellcolor[HTML]{FDE9E9}84.6 & \cellcolor[HTML]{FDE9E9}83.7 & \cellcolor[HTML]{FDE9E9}88.5 & \cellcolor[HTML]{FDE9E9}90.2 & \cellcolor[HTML]{FDE9E9}86.2 & \cellcolor[HTML]{FDE9E9}83.5\\   \bottomrule
\end{tabular}
\end{table*}

Table~\ref{tab:reference_model_results} shows the syntactic and functional correctness of the Python reference model, along with the line coverage of the final testbench generated in Stage 1 using GPT-4.. Since the results for GPT-3.5 are similar, we have chosen to omit them in the table. The experimental results demonstrate the effectiveness of generating reference models for various benchmarks, achieving nearly perfect syntactic correctness and over 90\% functional correctness across all benchmarks. Especially for the VerilogEval-machine benchmark, the reference model achieved an impressive functional correctness of 98.60\%. These findings demonstrate that LLMs are capable of generating a Python-based reference model with nearly correct functionality, even when the prompt describes hardware. Additionally, the line coverage of the final refined testbench remains consistently high across all benchmarks, with all coverage exceeding 90\%. The high coverage indicates that the generated test inputs effectively cover diverse functional scenarios of the hardware, making it effective in verifying the LLM-generated Verilog code.

\subsection{Functional Correctness of Verilog Code}

Following VerilogEval~\cite{VerilogEval}, we use the metric \textit{pass@k} to evaluate the functional correctness of the generated Verilog code. \textit{pass@k} represents the expected probability of at least one answer passing the functional evaluation when randomly choosing $k$ answers from $n$ candidates:

\begin{equation}
    \text{pass@k} := \mathbb{E}_{\text{Problems}} \left[ 1 - \frac{\binom{n-c}{k}}{\binom{n}{k}} \right]
\end{equation}
where $n$ represents the total number of generated samples for a given problem, $c$ denotes the number of correct samples, and $k$ specifies the number of selected code samples. We set $n=10$ according to the original settings.

We compared our approach with multiple categories of models: closed-source commercial LLMs (e.g., GPT-3.5~\cite{openai2023gpt35}, GPT-4~\cite{openai2023gpt4}, Claude3-Haiku~\cite{claude3_family}, and Claude3-Sonnet~\cite{claude3_family}), general open-source models (e.g., CodeLlama~\cite{roziere2023code}, CodeQwen~\cite{bai2023qwen}, and DeepSeek-Coder~\cite{guo2024deepseek}), and state-of-the-art Verilog-specific models (e.g., VerilogEval~\cite{VerilogEval}, CodeGen-6B MEV-LLM~\cite{nadimi2024multi}, BetterV-CodeQwen~\cite{BetterV}, RTLCoder~\cite{Xie_2023} and OriGen~\cite{2024origen}), which represent the latest advancements in research designed for hardware design tasks.

Table~\ref{tab:comparison} presents the functional correctness results across the benchmarks. Our approaches demonstrate significant improvements not only over leading commercial LLMs, such as GPT-4 and Claude3, but also domain-specific models, including OriGen and RTLCoder. Even when utilizing GPT-3.5 within our framework, we outperform most models. Although AutoVeriFix with GPT-3.5 shows slightly lower performance on the GPT-generated Verilog code (VerilogEval-machine) compared to OriGen, it significantly exceeds OriGen's performance on the more challenging VerilogEval-human benchmark and RTLLM. When employing GPT-4 within our framework, our approach achieves even greater performance, consistently outperforming all other models across all benchmarks.

In the RTLLM v2.0 benchmark, AutoVeriFix achieves remarkable performance, scoring 71.9 with GPT-3.5 and 83.5 with GPT-4, significantly outperforming all baseline models. For instance, compared to the original GPT-3.5, AutoVeriFix with GPT-3.5 improves the score on RTLLM v2.0 (pass@5) by 98.6\% (from 36.2 to 71.9). Moreover, on the VerilogEval-human benchmark, AutoVeriFix with GPT-4 achieves a pass@10 score of 84.6, outperforming both commercial model GPT-4 Turbo (72.4) and Claude3-Haiku model (60.9). The highest score recorded by AutoVeriFix was on the VerilogEval-machine benchmark, with a pass@10 score of 90.2, exceeding domain-specific models such as OriGen (85.7) and RTLCoder (81.8). This score is also quite close to the ideal performance of the reference model, as shown in Table~\ref{tab:reference_model_results}. Overall, these results indicate that AutoVeriFix with GPT-4 not only outperforms commercial general-purpose LLMs but also surpasses domain-specific models in generating Verilog code with functional correctness.

\subsection{Evaluation of Coverage Feedback}
\label{subsec:testeffect}


Upon reviewing the generated Verilog code, we found that all designs failing the testbench validation were functionally incorrect in the final evaluation. This highlights the filtering role of the testbench produced in the first stage as an early detection of faulty designs.

To enhance user confidence in LLM-based approaches for directly generating reliable Verilog and to minimize the need for additional modifications, we aim to ensure that all designs passing our testbench validation are functionally correct. However, achieving this expectation is challenging due to the incompleteness of the testbench, even with coverage feedback.

To evaluate the effectiveness of the testbench generated in the first stage, we define the false positive rate (FPR) for our approach:

\begin{equation} 
FPR = 1 - \frac{\text{\# designs that are functionally correct}}{\text{\# designs that pass the testbench}} 
\end{equation}

The FPR represents the percentage of designs that pass validation from the reference model's generated testbench but are functionally incorrect. A lower FPR indicates that the testbenches produced by the reference model are highly reliable, effectively filtering out invalid designs and ensuring that the majority of validated designs maintain high quality.

\begin{figure}
    \centering
    \includegraphics[width=0.95\linewidth]{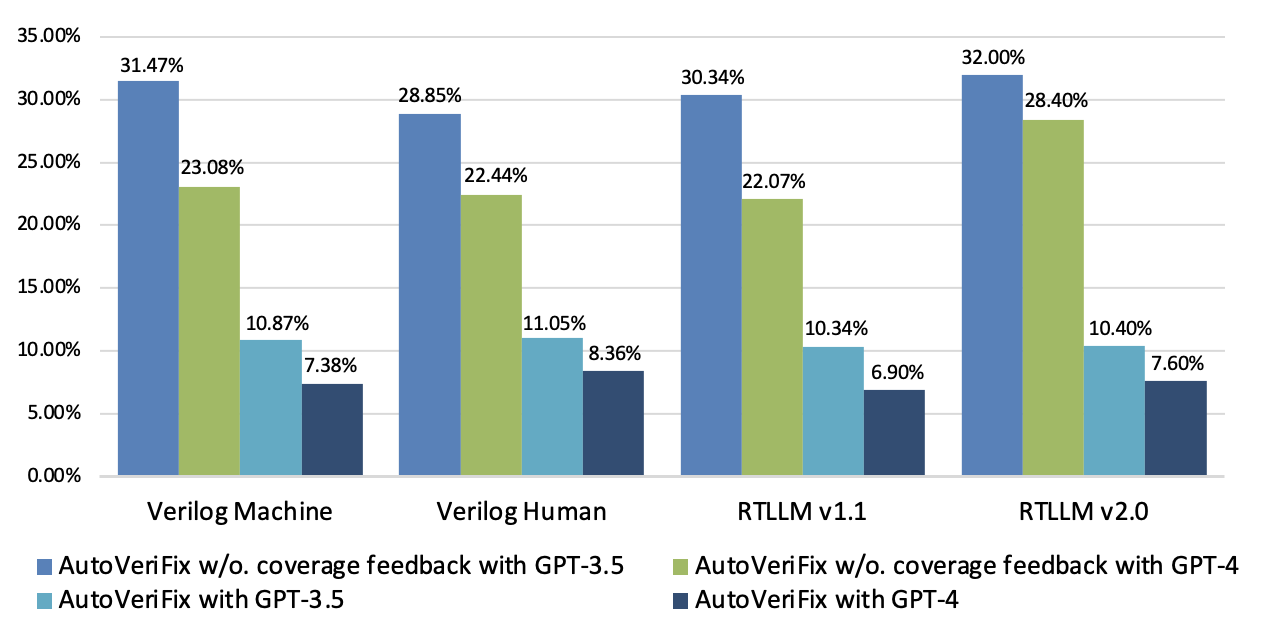}
    \caption{The percentage of designs that passed the validation of the testbench but are functionally incorrect.}
    \label{fig:falsepositive}
\end{figure}

The baseline for this experiment is the same framework illustrated in Fig.~\ref{fig:overview}, which consists of two stages: generating the reference model and producing the Verilog code, but without incorporating the coverage feedback prompt. Thus, this experiment compares the effectiveness of the initial testbench generated by the LLM in the first stage with the final testbench produced after applying coverage feedback. The comparison results, shown in Fig.~\ref{fig:falsepositive}, demonstrate that our method significantly reduces the FPR through the use of coverage feedback prompts.

Across all benchmarks, AutoVeriFix achieves an FPR of below 12\% with GPT-3.5 and below 9\% with GPT-4. These FPRs are considerably lower than those of the framework without coverage feedback, where GPT-3.5 has an FPR of approximately 30\% and GPT-4 ranges from 22\% to 29\%. For instance, in the case of using GPT-4 in RTLLM v2.0, the coverage feedback reduces the FPR from 28.4\% to 7.60\%. This substantial improvement underscores the effectiveness of our approach in filtering out low-quality Verilog code, ensuring that most generated Verilog designs passing the reference model's testbench validation are well-suited for direct use with minimal manual intervention.

\section{Conclusion}

This paper presents a two-stage framework for RTL code generation that harnesses the strengths of LLMs in Python code generation to create a Python-based reference model. In the first stage, the framework establishes Python reference models as functional standards, offering high-level implementations that guide the creation of testbenches. A feedback mechanism is incorporated, enabling the LLM to enhance testbench coverage by analyzing coverage metrics. In the second stage, the LLM iteratively addresses syntax errors with compiler feedback and corrects functional errors by identifying inconsistencies between the Verilog implementation and the testbench. These iterative improvements ensure that the generated Verilog RTL maintains a high level of functional correctness. Experimental results demonstrate that our approach outperforms both existing general-purpose LLMs and hardware-specific models in producing accurate Verilog code.

\bibliographystyle{IEEEtran}
\bibliography{references}

\begin{thebibliography}{10}
\providecommand{\url}[1]{#1}
\csname url@samestyle\endcsname
\providecommand{\newblock}{\relax}
\providecommand{\bibinfo}[2]{#2}
\providecommand{\BIBentrySTDinterwordspacing}{\spaceskip=0pt\relax}
\providecommand{\BIBentryALTinterwordstretchfactor}{4}
\providecommand{\BIBentryALTinterwordspacing}{\spaceskip=\fontdimen2\font plus
\BIBentryALTinterwordstretchfactor\fontdimen3\font minus \fontdimen4\font\relax}
\providecommand{\BIBforeignlanguage}[2]{{%
\expandafter\ifx\csname l@#1\endcsname\relax
\typeout{** WARNING: IEEEtran.bst: No hyphenation pattern has been}%
\typeout{** loaded for the language `#1'. Using the pattern for}%
\typeout{** the default language instead.}%
\else
\language=\csname l@#1\endcsname
\fi
#2}}
\providecommand{\BIBdecl}{\relax}
\BIBdecl

\bibitem{Mastropaolo_Pascarella_Guglielmi_Ciniselli_Scalabrino_Oliveto_Bavota_2023}
A.~Mastropaolo, L.~Pascarella, E.~Guglielmi, M.~Ciniselli, S.~Scalabrino, R.~Oliveto, and G.~Bavota, ``On the robustness of code generation techniques: An empirical study on github copilot,'' in \emph{2023 IEEE/ACM 45th International Conference on Software Engineering (ICSE)}.\hskip 1em plus 0.5em minus 0.4em\relax IEEE, 2023, pp. 2149--2160.

\bibitem{Nijkamp_Hayashi_Xiong_Savarese_Zhou}
E.~Nijkamp, H.~Hayashi, C.~Xiong, S.~Savarese, and Y.~Zhou, ``Codegen2: Lessons for training llms on programming and natural languages,'' \emph{arXiv preprint arXiv:2305.02309}, 2023.

\bibitem{Benchmark_RTL_2022}
S.~Thakur, B.~Ahmad, Z.~Fan, H.~Pearce, B.~Tan, R.~Karri, B.~Dolan-Gavitt, and S.~Garg, ``Benchmarking large language models for automated verilog rtl code generation,'' in \emph{2023 Design, Automation \& Test in Europe Conference \& Exhibition (DATE)}.\hskip 1em plus 0.5em minus 0.4em\relax IEEE, 2023, pp. 1--6.

\bibitem{Dehaerne_Verilog_2023}
E.~Dehaerne, B.~Dey, S.~Halder, and S.~De~Gendt, ``A deep learning framework for verilog autocompletion towards design and verification automation,'' \emph{arXiv preprint arXiv:2304.13840}, 2023.

\bibitem{OpenLLM}
S.~Liu, Y.~Lu, W.~Fang, M.~Li, and Z.~Xie, ``Openllm-rtl: Open dataset and benchmark for llm-aided design rtl generation(invited),'' in \emph{2024 IEEE/ACM International Conference on Computer-Aided Design (ICCAD)}.\hskip 1em plus 0.5em minus 0.4em\relax ACM, 2024.

\bibitem{VerilogEval}
M.~Liu, N.~Pinckney, B.~Khailany, and H.~Ren, ``Verilogeval: Evaluating large language models for verilog code generation,'' in \emph{2023 IEEE/ACM International Conference on Computer Aided Design (ICCAD)}.\hskip 1em plus 0.5em minus 0.4em\relax IEEE, 2023, pp. 1--8.

\bibitem{Xie_2023}
S.~Liu, W.~Fang, Y.~Lu, Q.~Zhang, H.~Zhang, and Z.~Xie, ``Rtlcoder: Outperforming gpt-3.5 in design rtl generation with our open-source dataset and lightweight solution,'' in \emph{2024 IEEE LLM Aided Design Workshop (LAD)}.\hskip 1em plus 0.5em minus 0.4em\relax IEEE, 2024, pp. 1--5.

\bibitem{2024origen}
F.~Cui, C.~Yin, K.~Zhou, Y.~Xiao, G.~Sun, Q.~Xu, Q.~Guo, D.~Song, D.~Lin, X.~Zhang \emph{et~al.}, ``Origen: Enhancing rtl code generation with code-to-code augmentation and self-reflection,'' \emph{arXiv preprint arXiv:2407.16237}, 2024.

\bibitem{chen2021codex}
M.~Chen, J.~Tworek, H.~Jun, Q.~Yuan, H.~P. D.~O. Pinto, J.~Kaplan, H.~Edwards, Y.~Burda, N.~Joseph, G.~Brockman \emph{et~al.}, ``Evaluating large language models trained on code,'' \emph{arXiv preprint arXiv:2107.03374}, 2021.

\bibitem{AutoChip}
S.~Thakur, J.~Blocklove, H.~Pearce, B.~Tan, S.~Garg, and R.~Karri, ``Autochip: Automating hdl generation using llm feedback,'' \emph{arXiv preprint arXiv:2311.04887}, 2023.

\bibitem{RTLFixer}
Y.~Tsai, M.~Liu, and H.~Ren, ``Rtlfixer: Automatically fixing rtl syntax errors with large language model,'' in \emph{Proceedings of the 61st ACM/IEEE Design Automation Conference}, 2024, pp. 1--6.

\bibitem{verilogcoder}
C.-T. Ho, H.~Ren, and B.~Khailany, ``Verilogcoder: Autonomous verilog coding agents with graph-based planning and abstract syntax tree (ast)-based waveform tracing tool,'' \emph{arXiv preprint arXiv:2408.08927}, 2024.

\bibitem{myhdl}
J.~Decaluwe, ``Myhdl: A python-based hardware description language,'' \url{http://www.myhdl.org}, 2003, accessed: July 2025.

\bibitem{soda}
Y.~Chi and J.~Cong, ``Exploiting computation reuse for stencil accelerators,'' in \emph{2020 57th ACM/IEEE Design Automation Conference (DAC)}.\hskip 1em plus 0.5em minus 0.4em\relax IEEE, 2020, pp. 1--6.

\bibitem{BetterV}
Z.~Pei, H.-L. Zhen, M.~Yuan, Y.~Huang, and B.~Yu, ``Betterv: Controlled verilog generation with discriminative guidance,'' \emph{arXiv preprint arXiv:2402.03375}, 2024.

\bibitem{thakur2024verigen}
S.~Thakur, B.~Ahmad, H.~Pearce, B.~Tan, B.~Dolan-Gavitt, R.~Karri, and S.~Garg, ``Verigen: A large language model for verilog code generation,'' \emph{ACM Transactions on Design Automation of Electronic Systems}, vol.~29, no.~3, pp. 1--31, 2024.

\bibitem{nijkamp2022codegen}
E.~Nijkamp, B.~Pang, H.~Hayashi, L.~Tu, H.~Wang, Y.~Zhou, S.~Savarese, and C.~Xiong, ``Codegen: An open large language model for code with multi-turn program synthesis,'' \emph{arXiv preprint arXiv:2203.13474}, 2022.

\bibitem{openai2023gpt35}
\BIBentryALTinterwordspacing
OpenAI, ``Gpt-3.5-turbo,'' 2023, accessed: 2024. [Online]. Available: \url{https://platform.openai.com/docs/models/gpt-3-5}
\BIBentrySTDinterwordspacing

\bibitem{openai2023gpt4}
\BIBentryALTinterwordspacing
------, ``Gpt-4 technical report,'' 2023, accessed: 2024. [Online]. Available: \url{https://openai.com/research/gpt-4}
\BIBentrySTDinterwordspacing

\bibitem{lu2024rtllm}
Y.~Lu, S.~Liu, Q.~Zhang, and Z.~Xie, ``Rtllm: An open-source benchmark for design rtl generation with large language model,'' in \emph{2024 29th Asia and South Pacific Design Automation Conference (ASP-DAC)}.\hskip 1em plus 0.5em minus 0.4em\relax IEEE, 2024, pp. 722--727.

\bibitem{nadimi2024multi}
B.~Nadimi and H.~Zheng, ``A multi-expert large language model architecture for verilog code generation,'' in \emph{2024 IEEE LLM Aided Design Workshop (LAD)}.\hskip 1em plus 0.5em minus 0.4em\relax IEEE, 2024, pp. 1--5.

\bibitem{roziere2023code}
B.~Roziere, J.~Gehring, F.~Gloeckle, S.~Sootla, I.~Gat, X.~E. Tan, Y.~Adi, J.~Liu, R.~Sauvestre, T.~Remez \emph{et~al.}, ``Code llama: Open foundation models for code,'' \emph{arXiv preprint arXiv:2308.12950}, 2023.

\bibitem{bai2023qwen}
J.~Bai, S.~Bai, Y.~Chu, Z.~Cui, K.~Dang, X.~Deng, Y.~Fan, W.~Ge, Y.~Han, F.~Huang \emph{et~al.}, ``Qwen technical report,'' \emph{arXiv preprint arXiv:2309.16609}, 2023.

\bibitem{guo2024deepseek}
D.~Guo, Q.~Zhu, D.~Yang, Z.~Xie, K.~Dong, W.~Zhang, G.~Chen, X.~Bi, Y.~Wu, Y.~Li \emph{et~al.}, ``Deepseek-coder: When the large language model meets programming--the rise of code intelligence,'' \emph{arXiv preprint arXiv:2401.14196}, 2024.

\bibitem{claude3_family}
{Anthropic}, ``Introducing the next generation of claude,'' \url{https://www.anthropic.com/}, 2024.

\end{thebibliography}

\end{document}